\documentclass{aa}
\usepackage{epsfig}
\usepackage{natbib}
\bibpunct{(}{)}{;}{a}{}{,}

\begin{document}

%\fraction makes a nice fraction
\def\fraction#1/#2{\leavevmode\kern.1em
 \raise.5ex\hbox{\the\scriptfont0 #1}\kern-.1em
 /\kern-.15em\lower.25ex\hbox{\the\scriptfont0 #2}}
%\simlt and \simgt produce > and < signs with twiddle underneath
\def\simlt{\lower.5ex\hbox{$\; \buildrel < \over \sim \;$}}
\def\simgt{\lower.5ex\hbox{$\; \buildrel > \over \sim \;$}}
\def\etal{et al.\ }

\title{Observations of the High-redshift Galaxy B2 0902+343 at 92 cm}
\author{Ann Marie Cody\inst{1,2} \and
Robert Braun\inst{1}
}

\institute{
Netherlands Foundation for Research in Astronomy,
    P.O. Box 2,
    7990 AA Dwingeloo,
    The Netherlands \and
Dept. of Astronomy, Harvard University, 60 Garden St., Cambridge, 
MA 02138, U.S.A.
%\and
}
\offprints{A.M. Cody \\
\email{acody@cfa.harvard.edu}}
\date{Received date / Accepted date}

\abstract{ We present 92~cm observations of the high-redshift galaxy
B2~0902+343 (z=3.40) with the Westerbork Synthesis Radio Telescope
(WSRT). Twenty-one~cm \ion{H}{i} is detected in absorption, at a depth
of $15.9$~mJy (=~$6\sigma$) and rest frame velocity width
120~km~s$^{-1}$. We also report a null detection of OH emission or
absorption at the 1665/1667~MHz rest-frame transitions.  Based on our
spectral sensitivity, we derive an upper limit on OH maser luminosity
of $\log(L/L_{\sun})<5.2$. In addition, we consider the possibility of
absorption, and estimate a maximum column depth N(OH) of
$1\times10^{15}~\mathrm{cm}^{-2}$, corresponding to at most
$2\times10^5-9\times10^6\mathrm{M}_{\sun}$ of OH in the galaxy. The
implications of these results are discussed in the context of current
efforts to extend the OH megamaser luminosity function and galactic
merger rate to high redshift.  
}

\maketitle

\section{Introduction}

  Astronomical interest in the early universe, and in particular the epoch
of galaxy formation, has spurred numerous attempts to detect and analyze
high-redshift objects.  Much attention has recently been devoted to the
study of galaxies at z~$>1$, as these sources are thought to serve as
probes into gas and dust environments at early times.  At $z$=3.40
\citep{1988ApJ...333..161L,1991PhRvL..67.3328U}, the radio source
\object{B2 0902+343} (second Bologna survey, \citealt{1972A&AS....7....1C})
possesses one of the highest known redshifts among radio galaxies.  With
the exception of quasars, relatively few such distant objects are known.
 
  Both optical and high-resolution radio observations of \object{B2
0902+343} have turned up rather peculiar structures.  
\citet{1988ApJ...333..161L} first reported \object{B2 0902+343} to be a
galaxy, noting that highly redshifted optical Lyman-$\alpha$ emission is
nearly coincident with the radio source.  \citet{1999A&A...341..329P}
detected two optical components with the Hubble Space Telescope, and
observed that they are not aligned with distinct lobes in the radio
continuum at 1.65~GHz. Other complex radio features were seen by
\citet{1994AJ....107.2299C} and \citet{1995A&A...298...77C}. CO lines, on
the other hand, have not been detected \citep{1996A&A...313...91D}
Nevertheless, the existence of gas \emph{is} implied by the significant
\ion{H}{i} absorption feature at 323~MHz first observed by
\citet{1991PhRvL..67.3328U} and later confirmed by
\citet{1993ApJ...415L..99B} and \citet{1996CGHR.......171S}, who noted a
potentially significant absorption wing to the blueward side of this line.  
\object{B2~0902+343} happens to be one of only two galaxies at z~$>3$ that
exhibit an \ion{H}{i} absorption feature. It is still unclear which of the
components seen in high-resolution radio images is (partially) obscured
by foreground gas.

  In the quest for further insight into the molecular contents of
\object{B2~0902+343}, we have tested the possibility that it may host an OH
megamaser. The companion OH lines at 1665 and 1667~MHz are difficult to
observe at moderate redshift because of interference from terrestial
television transmission and a shortage of suitably equipped telescopes.  
Beyond $z$=3.3, however, OH once again becomes accessible. As
\citet{2001MNRAS.328L..17T}, \citet{1998A&A...336..815B}, and others have
suggested, there is a strong chance that most ultraluminous and
hyperluminous infrared galaxies (ULIRGs and HLIRGs), as well as their
high-redshift submillimeter counterparts at $z$=1--10, display associated
OH maser emission.  At lower redshifts, a survey performed with the Arecibo
Telescope by \citet{2000AJ....119.3003D, 2001AJ....121.1278D,
2002AJ....124..100D,2002ApJ...572..810D} has already turned up 50 new OH
megamasers at $z$=0.1--0.2.  Bolstering the possibility of OH maser
activity at redshift 3.4 in \object{B2~0902+343} are the presence of
neutral hydrogen absorption and strong continuum emission from this galaxy,
key ingredients in the pumping of megamaser sources.

  With the existing spectral observations and potential OH detection in
mind, we have undertaken a series of observations with the recently
upgraded Westerbork Synthesis Radio Telescope (WSRT), at the 92~cm
(310--390~MHz) band. This spectral region is ideally placed for
observation of both \ion{H}{i} and OH in \object{B2~0902+34}, since they
fall at frequencies near 323~MHz and 379~MHz, for a redshift $z$=3.40.
The present work has set out to clarify the structure of
\object{B2~0902+343}'s neutral hydrogen spectrum and to search for the
presence of OH in the galaxy.  In the following discussions, we
describe our analysis and the resulting spectra.

\section{Observations and Data Reduction}
\begin{figure*} 
\begin{center}  
    \resizebox{18cm}{!}{\includegraphics{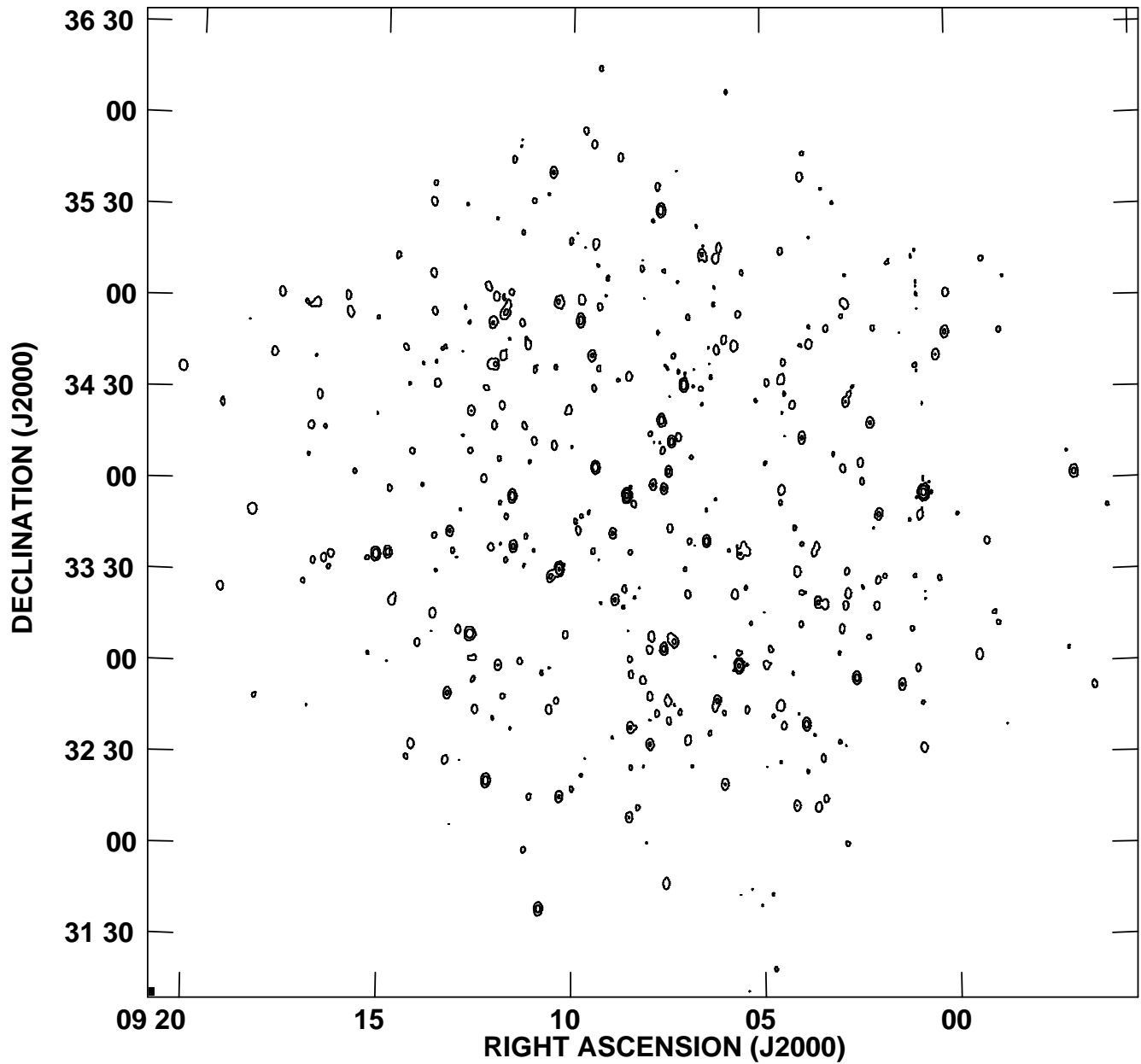}}
\end{center}  
\caption{The field around \object{B2~0902+343} is littered with continuum
sources at 92~cm.  Contours are drawn at levels of -5, 5, 50
and 500 mJy/beam. The extent of the primary beam can be seen as the
central region of this clean map, where the most sources are detected.}
\label{fig1}
\end{figure*}

  Observations of \object{B2~0902+343} were made on three separate
occasions with the WSRT.  The interferometer consists of fourteen 25~m
dishes arranged in an east-west configuration; baselines range from
36~m to 2.7~km.  The 25~m telescopes offer a primary beam of about
2.6\degr~FWHM in the 92~cm band. All observations were carried out with
a total bandwidth 5~MHz, centered near the frequencies of the OH and
\ion{H}{i} lines.  In the 92~cm band, the system
temperature is approximately 125~K, giving an expected sensitivity of
0.35 mJy/beam for 12-hour observations.  The source
confusion limit at 92~cm is just below this level, at 0.3~mJy/beam.  A map of
continuum sources near \object{B2~0902+343} is displayed in Fig.\
\ref{fig1}.  Not only are observations in the 92~cm band complicated by
contaminating emission from numerous unresolved continuum sources, but
they are also affected by widespread radio frequency interference (RFI)
generated by both local equipment and TV broadcasts.  Interference had to
be properly removed from the data, and continuum sidelobe structure
subtracted before useful images and spectra could be made.

\subsection{1998 Observations} 

  Two WSRT observations to search for OH in \object{B2~0902+343} were
carried out in the 1998; these took place on October 30 and November 14.  
The synthesized beam measured $100\arcsec\times 60\arcsec$, and data were
taken in 256 spectral channels, in 19.5~kHz increments centered about
379~MHz. Integration times for the source were 3 hours and 12 hours,
yielding a total expected sensitivity of approximately 8~mJy per channel.  
On the first date, the quasar 3C~286 was observed for one hour to acquire
flux calibration data. During the second run, both 3C~48 and 3C~286 were
observed for the same purpose. All data reduction was carried out with the
NRAO AIPS package.  Based on \citet{1977A&A....61...99B}, flux densities of
24.7~Jy were calculated for 3C~286 and 39.6~Jy for 3C~48. At the
wavelengths in question, \object{B2~0902+343} is an unresolved source of
about 1~Jy.  After amplitude, phase, and bandpass calibrations were carried
out separately for each observation, both were merged into a single data
set.

  Continuum flux models were produced by running the CLEAN iterative source
deconvolution algorithm on a ``continuum'' u-v data set composed of
visibilities averaged over the central 75\% of the band.  Although some of
the central channels may contain spectral features, toward
\object{B2~0902+343} they will be substantially diluted over a 3.75~MHz
bandwidth, as well as over the many other bright continuum sources in the
field.  The CLEAN continuum images were subsequently employed in a
self-calibration routine, and the resulting complex gain solutions were
applied to all channels.  Broad-band emission was substracted from the
data, and channel maps were then constructed to verify that there were no
remaining point sources or stray sidelobes near the position of
\object{B2~0902+343}.  
\begin{center}
\begin{table}
\begin{tabular}{lcc}
\hline
\hline
Parameter & 1998 Observations & 2002 Observation\\
\hline
Frequency center(s)  & 379~MHz & 323 and 379~MHz \\
Bandwidth            & 5 MHz   & 5 MHz \\
Number of Channels   & 256     & 512 \\
Number of IFs        & 1       & 2 \\
Number of receivers & 11      & 13 \\
On-source time       & 15 hrs. & 12 hrs. \\
Expected sensitivity & 6.5 mJy & 8.6 mJy \\
  per channel  &         & \\
\hline
\end{tabular}
\caption{Observational Parameters}
\label{tab1}
\end{table}
\end{center}
\subsection{2002 Observation} 
  The most recent observation WSRT of \object{B2~0902+343} was carried out
on March 15, 2002.  The backend provided two independent IFs, tuned to
379~MHz and 323~MHz for OH and HI. The number of channels was increased to
512, offering higher spectral resolution in frequency increments of
9.8~kHz, and the synthesized beam measured $97\arcsec \times 56\arcsec$.
\object{B2~0902+343} was observed for a total of 12 hours, yielding an
expected sensitivity per channel of $\sim6$~mJy, consistent with our
measured RMS values.  Additional pointings were devoted to 3C147 and 3C295
for flux scale calibration; flux densities of 49.0~Jy and 55.6~Jy,
respectively, were adopted for these sources based on the
\citet{1977A&A....61...99B} formula. Amplitude, phase, and frequency
calibrations were carried out with standard AIPS tasks.  CLEANing was
performed on continuum data sets constructed from only 40 central spectral
channels; averages over larger frequency ranges left unwanted residual
sidelobes for off-axis sources within the primary beam due to bandwidth
smearing. Due to the array's east-west configuration, and hence coplanar
u-v points, sidelobe response from bright sources far outside the primary
beam also proved to be a problem in the \ion{H}{i} IF, even after extensive
CLEANing, self-calibration, and continuum subtraction. However, the
channels in which the most prominent sidelobe feature crosses the
\object{B2~0902+343} position are far enough from the 323~MHz \ion{H}{i}
frequency to ensure that the absorption line and any broad wing as proposed
by \citet{1996CGHR.......171S} remain uncontaminated.  Observational
parameters for both the 1998 and 2002 sessions are summarized in Table
\ref{tab1}.

\subsection{Spectra}
\begin{figure} 
\begin{center}  
%\vspace*{0.0cm} 
%\hspace*{-3.0cm}
\epsfig{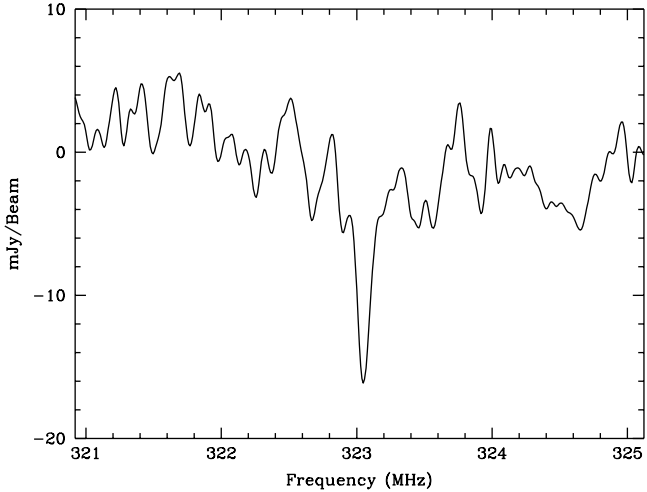}   
%\vspace{-2.0cm} 
\caption{The \object{B2~0902+343} spectrum in a 5~MHz bandwidth around
323~MHz.  The spectrum has been smoothed to a resolution of 60.5~kHz.
\ion{H}{i} absorption is clearly visible at a frequency just above
323~MHz.}
\label{fig2}
\end{center}
\end{figure}

  Given the noisy appearance of the spectra, we chose to smooth them to a
resolution approximately 1/3 of the previously observed \ion{H}{i}
linewidth in \object{B2~0902+343}.  Therefore, the 1998 OH-frequency
spectrum was filtered with a Gaussian FWHM of 3.1 channels, while the
same was done for the 2002 OH and \ion{H}{i} spectra with 6.2 channels,
due to the increased spectral resolution of the more recent
observations. To create the highest possible signal-to-noise ratio for
an OH detection, we performed a weighted average of the two separate OH
spectra.  Weights were based on the RMS noise levels measured in each
(3.2--3.5~mJy), and the RMS of the final smoothed spectrum is
2.4~mJy/beam at $\Delta \nu$=60.5~kHz. For the single \ion{H}{i}
spectrum, the RMS is nearly the same, at 2.5~mJy/beam. Rough estimates
of the continuum levels were made by averaging flux densities across
the spectra, leaving out the obvious absorption region in the case of
HI.  No slope was allowed for in the continuum, since the both spectra
appear fairly flat and far-source sidelobe contamination in the
\ion{H}{i} spectrum may have produced a disruptive bump over a small
frequency region. The spectra are presented in Figs.\ \ref{fig2} and
\ref{fig3}.

\section{Discussion}

\subsection{\ion{H}{i} Absorption} 

  A narrow \ion{H}{i} absorption line is clearly detected at 323.053~MHz,
with a depth of $15.9$~mJy (=~$6\sigma$), relative to the 1.2~Jy total
flux, and a Gaussian FWHM of $0.13\pm0.05$~MHz. This value corresponds
to a rest frame velocity width of 120~km~s$^{-1}$.  Our \ion{H}{i}
linewidth is in good agreement with the values of 90--100~km~s$^{-1}$
exhibited in the \citet{1993ApJ...415L..99B} and
\citet{1996CGHR.......171S} spectra.  Thus, the implied values for
column density and total mass of \ion{H}{i} absorbing material are
similar to those inferred by \citet{1996CGHR.......171S}:
$N(HI)\sim3\times10^{21}\mathrm{cm}^{-2}$ assuming a mean spin
temperature of 1000~K, and
M$_\mathrm{\ion{H}{i}}=10^{7}-10^{10}~\mathrm {M}_{\sun}$, depending
on whether the gas at this column density is assumed to cover a region
of only 1.5~kpc$^2$, corresponding to the radio continuum hotspot, or as
much as 750~kpc$^2$, corresponding to the entire Ly-$\alpha$ emission
region. The possible absorption feature detected by
\citet{1993ApJ...415L..99B} near 322.5~MHz may be present in our
spectrum, but noise levels preclude any confident identification.  The
same is true of the broad wing component detected by
\citet{1996CGHR.......171S} on the blueward side of the \ion{H}{i}
line. 

\begin{figure}
\begin{center}
%\vspace*{0.3cm}
%\hspace*{-6.0cm}
\epsfig{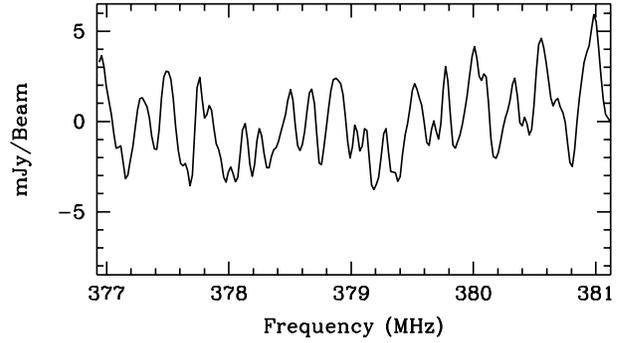}
%\vspace*{-7.0cm}
\caption{The \object{B2~0902+343} spectrum in a 5~MHz bandwidth around
379~MHz. The spectrum is smoothed to 60.5~kHz and reveals no sign of OH
emission or absorption near 379~MHz.  The RMS noise is approximately
2.5~mJy.}
\label{fig3}
\end{center}
\end{figure}

\subsection{Non-detection of OH Emission} 

  Although the signature of \ion{H}{i} at 323~MHz is clear in a 12-hour
observation, we fail to detect any OH lines in the combined
377--381~MHz band spectrum. As seen in Fig.\ \ref{fig3}, it appears
exceedingly flat. With a measured RMS of 2.4~mJy at channel intervals
of $\Delta \nu=60.5~$kHz, we can set an upper limit of
$\sim$1.5$\sigma$=3.6~mJy for the extent of any undetected spectral
line features.

  Due to lack of data points, the OH megamaser luminosity function
\citep[c.f.][]{1998A&A...336..815B,2002ApJ...572..810D} is poorly
constrained at high redshift.  Therefore, while there is little evidence to
point toward the existence of hydroxyl gas in \object{B2~0902+343}, it is
nevertheless instructive to determine how luminous the system could be in
OH emission before exceeding the detection limits.  Using a formulation
analogous to \citet{2002AJ....124..100D, 2002ApJ...572..810D}, we calculate
the maximum OH luminosity based on a 1.5$\sigma$ detection limit and an
average rest frame spectral line width of $\Delta v$~=~150~km~s$^{-1}$. The
maximum OH luminosity is then L=$4\pi\times D_{L}^{2}\times
f_\mathrm{obs}$, where $f_\mathrm{obs}=(1.5\sigma)(\nu_{o}\Delta
v)(1+z)^{-1}c^{-1}$ is the integrated emission line flux in the observer's
frame, and $D_{L}$ is the luminosity distance.  
Combining this value with the most recent cosmological paraemters
$\Omega_\mathrm{m}=0.3$, $\Omega_{\Lambda}=0.7$,
H$_{o}=75$~km~s$^{-1}~\mathrm{Mpc}^{-1}$ yields $D_{L}=$27,520~Mpc.
Applying the luminosity equation, this translates to a maximum
$\log(\frac{L}{L_{\sun}})=5.2$. For other cosmologies, the value is found
to be around 5.

  Because of the high redshift of \object{B2~0902+343}, the maximum
possible OH luminosity for this galaxy is quite large. \citet{
2001MNRAS.328L..17T} and \citet{1998A&A...336..815B} have already pointed
out the limitations that detector sensitivity places on ability to observe
all but perhaps the brightest OH megamasers in high-redshift sources. In
the case of \object{B2~0902+343}, the calculated maximum luminosity serves
as a lower limit on detectable OH luminosity imposed by our particular
observing parameters. Comparing this limit against luminosities of
\emph{known} OH megamasers in the \citet{2002ApJ...572..810D} survey at
$z$=0.1--0.25, we find that no OH megamasers more powerful than our
detection threshold have been discovered thus far.

  However, there is still reason to believe that more luminous masers
exist at high redshift.  Strong correlation is predicted to exist
between galaxy merger rate \citep[(1+z)$^{m}$,][]{1998A&A...336..815B,
2002AJ....124..100D,2002ApJ...572..810D, 1996Ap....39...327}, far
infrared luminosity ($L_\mathrm {FIR}$), and OH megamaser rate
\citep[][and references
therein]{2001MNRAS.328L..17T,1998A&A...336..815B}, although it has yet
to be verified for objects at high redshift.
\citet{2002AJ....124..100D} have estimated linear relations between OH
and far infrared luminosity based on data from known OH megamasers and
find a best fit of log($L_\mathrm{OH})=(1.2\pm0.1)\log(L_\mathrm
{FIR})-(11.7\pm1.2)$. The far infrared luminosity corresponding to our
OH luminosity upper limit is then L$_\mathrm{FIR}\sim 10^{14}$, which is in
the ``hyperluminous" regime.  The \object{B2~0902+343} data point
provides a preliminary constraint on the number of such galaxies at
$z\simgt 3$.

\subsection{Possibility of OH Absorption}

  Although \object{B2~0902+343} may not exhibit detectable maser lines, we
note that the absence of emission does not necessarily imply the absence of
OH gas.  If emission is to be observed in the 1665/1667~MHz transitions,
the OH molecules must lie along our line of sight, be pumped by an
appropriate far-IR source and lie in front of a bright radio continuum
source. These restrictive secondary conditions are only met some fraction
of the time.  Similar arguments may be made for an OH \emph{absorption}
scenario.  It is plausible that OH is present in our observation throughout
a moderate range of velocities, yet with a small enough opacity or covering
factor that absorption can not be detected above the noise floor.  Since
this is clearly not the case for HI, any OH gas which is present is likely
to have a significantly lower absorption optical depth or be spatially
offset from the hydrogen absorbers. That may be true if most of the
\ion{H}{i} gas is located in the continuum ``hot spot'' observed by
\citet{1995A&A...298...77C}, while OH is instead associated with other
regions of the galaxy.

  On the whole, OH absorption is probably of less interest than emission,
since it is unlikely to shed light on the galaxy merger rate. However,
studies of absorption could offer insight into the abundance of molecular
gas in the early universe.  The signature of OH absorption has already been
detected in the spectra of some low to moderate redshift galaxies,
including \object{PKS~1830-21} at $z=0.89$ \citep{1999A&A...343L..79C}. In
the case of \object{B2~0902+343}, we derive an upper limit for OH column
density and mass.  With an RMS limit of 2.4~mJy and a total continuum flux
of 1.1~Jy, the optical depth, $\tau$, for any absorption line is less than
$\sim 0.3\%$. We adopt a linewidth of 120~km~s$^{-1}$, based on value
measured for \ion{H}{i} in \object{B2~0902+343}. From Liszt \& Lucas's
measurements of galactic OH absorption and emission, we select
T$_\mathrm{ex}=T_\mathrm{CMB}+1$ of order 10~K.  Together, these parameters
give a maximum N(OH) of $8\times0^{14}~{\rm cm}^{-2}$.  A rough check of
this value may be carried out by considering the ratio
N(OH)/N(HI)$<5\times10^{-8}$ for galactic sources that follows from
\citet{1996A&A...314..917L} and \citet{1996A&A...307..237L}.  With N(HI)
approximately $3\times10^{21}$ (assuming a mean HI spin temperature of 1000
K), we expect an OH column density less than
$1.5\times10^{14}$~atoms~cm$^{-2}$. The implication is that our upper limit
to OH absorption is roughly consistent with the Galactic ratio of
N(OH)/N(HI).

  To convert from column density to molecular mass, it is necessary to
specify the amount of cross-sectional area that OH molecules might occupy
in \object{B2~0902+343}. At $z=3.40$, these angular sizes are a function of
angular diameter distance and linear size.  One arcsecond at the
distance of \object{B2~0902+343} corresponds to 6.9~kpc ($\Omega_{m}=0.3$,
$\Omega_{\Lambda}=0.7$, $H_{o}=75~\mathrm{km s}^{-1}\mathrm{Mpc})$, or
values in the neighborhood of 5--10~kpc for other cosmologies.  Adopting
the same range of areas considered previously for the HI absorbing region
of only 1.5~kpc$^2$, corresponding to the radio continuum hotspot or as
much as 750~kpc$^2$, corresponding to the entire Ly-$\alpha$ emission
region, yields OH mass limits of between $2\times10^5\mathrm{M}_{\sun}$ and
$9\times10^6\mathrm{M}_{\sun}$

\section{Summary} 
  With the WSRT, we have achieved a redetection of HI 21~cm absorption in
B2~0902+343 at z~=~3.3962, at a significance in excess of $6\sigma$. Our
linewidth and depth are consistent with previous observations.  Our
non-detection of associated redshifted OH 1665/1667 emission or absorption
at the 0.2\% level yields an (1.5$\sigma$) upper limit to the OH mega-maser
luminosity of $\log(\frac{L}{L_{\sun}})=5.2$, and an upper limit to the OH
column density along this line-of-sight, N(OH) $< 10^{15}~{\rm cm}^{-2}$.  
Although no OH emission or absorption is detected, our upper bounds on
these values may help to constrain predictions of gas quantity and merger
activity at high redshift, once a larger galaxy sample is available. We
encourage work to search for OH megamasers in other high redshift galaxies
at $z>3$.  Future observations with more sensitive instruments, such as the
proposed Square Kilometer Array \citep[SKA;][]{2001AAS...199.7302B}, will
provide for either new detections or further limits. Higher angular
resolution could also pinpoint the location of the HI gas in relation to
the optical, infrared, and radio continuum emission features.

\begin{acknowledgements}
A.M.C. is grateful for the support of a summer research grant sponsored
by the Netherlands Foundation for Research in Astronomy. The Westerbork
Synthesis Radio Telescope is operated by the Netherlands Foundation for
Research in Astronomy under contract with the Netherlands Organization
for Scientific Research.
\end{acknowledgements}

\bibliographystyle{apj}
\bibliography{0902}

\end{document}